\documentclass[aps,prc,twoside,nofootinbib,showpacs]{revtex4}
\usepackage{amsmath,amssymb}
\usepackage{graphicx}
\begin{document}

\title{\boldmath Coherent $\pi^0\pi^0$ photoproduction on lightest nuclei}
\author{M. Egorov}
\affiliation{Laboratory of Mathematical Physics, Tomsk Polytechnic University, 634050
Tomsk, Russia}
\author{A. Fix}\email{fix@tpu.ru}
\affiliation{Laboratory of Mathematical Physics, Tomsk Polytechnic University, 634050
Tomsk, Russia}
\date{today}

\begin{abstract}
Coherent photoproduction of $\pi^0\pi^0$ on the deuteron and $^3$He is calculated. The
isoscalar and isovector parts of the elementary photoproduction amplitude was determined
by fitting the measured total cross section on protons and neutrons. The dependence of
the cross section on the isospin of the target nucleus is discussed.
\end{abstract}


\maketitle


%
\section{Introduction}
Coherent photoproduction of neutral pions on nuclei, when the target nucleus remains in
its ground state, are widely used in meson-nuclear physics. Firstly, these reactions
allow obtaining nuclear structure information. Since the incident photon probes the
entire nuclear volume, the cross section directly depends on the nuclear density
distribution. An important advantage of reactions involving neutral pions is that they
are not complicated by the coulomb interaction, so that the description of the final
state becomes more simple. Among the recent works one can mention
Ref.\,\cite{KruscheLi7}, in which the dependence of the cross section for
$\gamma\,^7$Li$\to\pi^0\,^7$Li on the $^7$Li parameters was explored. Generally, the
results obtained from the photoproduction reactions seem to be of comparable quality to
that provided by other processes, for example, by elastic particle-nuclear scattering.

The second important aspect, related to the coherent meson photoproduction, is the
investigation of the photoproduction mechanism. The case in point are the spin-isospin
selection rules provided by the quantum numbers of the target nucleus. These rules  make
it possible to identify the contributions of individual components of the photoproduction
amplitude. Recently, experiments were carried out in order to study partial transitions
in nuclei proceeding when the meson is produced \cite{Watts}. Such experiments became
possible due to significant improvements in detection of the mesons. This allowed one to
overcome difficulties in identifying individual nuclear transitions, which were the main
obstacle of using nuclear reactions to study meson photoproduction dynamics.

In the present work we consider coherent photoproduction of two neutral pions
\begin{equation}\label{10}
\gamma+A\to\pi^0+\pi^0 +A\,.
\end{equation}
The corresponding single nucleon process
\begin{equation}\label{11}
\gamma+N\to\pi^0+\pi^0 +N
\end{equation}
is still not very well understood. There are significant deviations between the analyses
provided by different groups (see, for example, the discussion in Ref.\,\cite{KaFiPra}).
As a rule they give an acceptable description of the available data, although the partial
wave content of the corresponding amplitudes may be rather different. The complete
experiment, which may resolve this ambiguity, appears to be extremely difficult when two
mesons are produced \cite{FiArPW}. Some results for photoproduction of $\pi^0\pi^0$ on
protons are obtained only in the low energy region \cite{KaFiPra} where one can
effectively use a truncated expansion of the amplitude over partial waves.

The information which may be obtained on the elementary amplitude is usually considered
as a main motivation for studying coherent photoproduction on nuclei. There are, however,
difficulties arising from a rather strong sensitivity of the coherent cross section to
different model ingredients, like off-shell behavior of the elementary amplitude, effects
of the final pion-nuclear interaction, role of two-nucleon production mechanisms, {\it
etc}. These problems seem to be unavoidable at least within the models in which the
impulse approximation is used as a basic zero-order approximation. Hence, more or less
firm quantitative conclusion can be drawn only as long as sophisticated microscopic
models are adopted. At the same time, in the $\pi^0\pi^0$ case where even the general
partial wave structure is rather poorly known, already qualitative results may provide
very useful information.

Here we are mainly focused on the isospin structure of the $\pi^0\pi^0$ photoproduction
amplitude. The corresponding selection rules provided by the charge of the nucleus enable
us to get information about the role of isoscalar and isovector transitions in this
reaction. The numerical calculations are performed on the deuteron and $^3$He. Our main
purpose is to study the presumably strong sensitivity of the cross section to the isospin
of the target. The results are expected to be important as a firm testing ground for our
knowledge on the two-pion photoproduction dynamics.

\section{Formalism}

The formalism, which we used to calculate the cross section (\ref{10}), was partially
considered in our previous work \cite{Egorov13} on coherent photoproduction of
$\pi^0\eta$ pairs. The cross section for coherent photoproduction of two mesons on a
nucleus $A$ with spin $J$ is given in the center-of-mass frame by
\begin{equation}\label{15}
\frac{d\sigma}{d\Omega_p d\Omega_{q^*}
d\omega_{\pi\pi}}=\frac{1}{(2\pi)^5}\frac{E_AE_A^\prime q^*p}{8E_\gamma
(E_\gamma+E_A)^2}\frac{1}{2(2J+1)}\sum\limits_{\lambda
M_iM_f}\left|\,T^\lambda_{M_iM_f}\right|^2\,,
\end{equation}
where the matrix element $T^\lambda_{M_iM_f}=\langle M_f|\widehat{T}^\lambda_{\gamma
A}|M_i\rangle$ determines the transition between the nuclear states with definite spin
projection. The energy of the incident photon and the initial and the final energies of
the target nucleus are denoted respectively by $E_\gamma$, $E_A$ and $E_A^\prime$. As
independent kinematical variables we took the invariant $\pi\pi$ mass $\omega_{\pi\pi}$,
the spherical angle $\Omega_p$ of the final nucleus momentum $\vec{p}=(p,\Omega_p)$ and
the spherical angle of the momentum $\vec{q}^{\,*}=(q^*,\Omega_{q^*})$ of one of the two
pions in their center-of-mass frame. The photon polarization index $\lambda=\pm 1$ will
be omitted in subsequent expressions.

\begin{figure}
\begin{center}
\resizebox{0.8\textwidth}{!}{%
\includegraphics{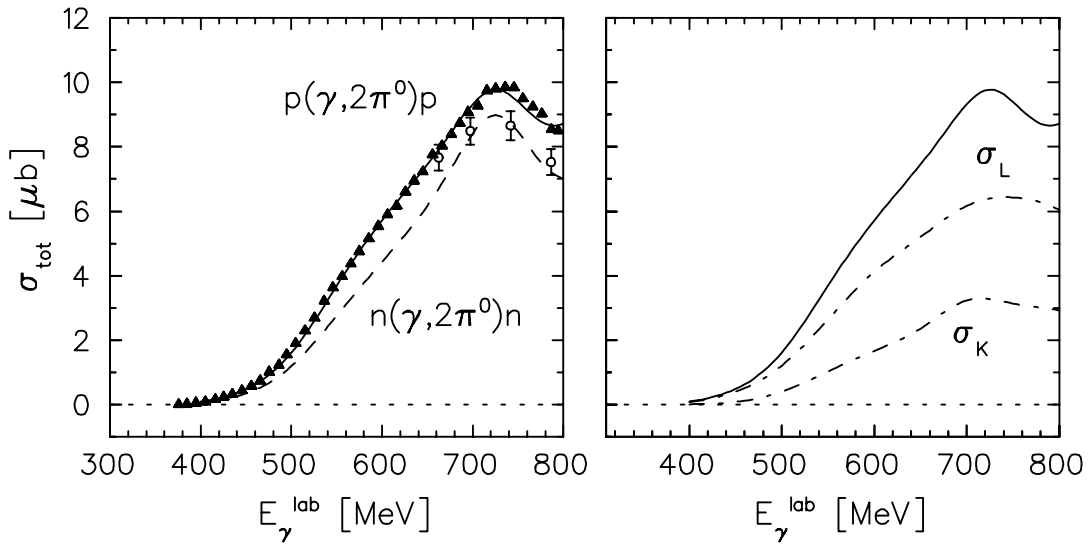}
} \caption{Left panel: total cross section for $\pi^0\pi^0$ photoproduction on protons
(solid line) and neutrons (dashed line). The data are taken from
Ref.\,\protect\cite{KaFiPra} (triangles) and \protect\cite{Ajaka2pi0n} (circles). Right
panel: solid line: total cross section for $\pi^0\pi^0$ photoproduction on protons (same
as on the left panel), dash-dotted curves: the proton cross section calculated with only
the spin-flip $\vec{L}$ and the spin independent part $K$ in Eq.\,(\ref{24}).}
\label{fig1}
\end{center}
\end{figure}

In order to calculate the amplitude $T_{M_iM_f}$ we used the impulse approximation taking
the operator $\widehat{T}_{\gamma A}$ as a superposition of operators
$\hat{t}_{\gamma N}$ for photoproduction on free nucleons
\begin{equation}\label{18}
\widehat{T}_{\gamma A}=\sum\limits_i\hat{t}_{\gamma N}(i)\,,
\end{equation}
where the summation is performed over all nucleons in the target. The spin structure of
the single nucleon operator has the well known form
\begin{equation}\label{24}
\hat{t}_{\gamma N}=K+i\vec{L}\cdot\vec{\sigma}\,,
\end{equation}
where $\vec{\sigma}$ is the Pauli spin operator, and the amplitudes $\vec{L}$ and $K$
describe transitions with and without flip of the nucleon spin. Both amplitudes depend on
the photon polarization index $\lambda$ and on the kinematical variables as described
above.

For
$\hat{t}_{\gamma N}$ we adopted the phenomenological analysis \cite{KaFiPra} for $\gamma
p\to\pi^0\pi^0 p$. The analysis contains the resonance $D_{13}(1520)$ whose parameters
were taken from the Particle Data Group listing (PDG) \cite{PDG} and were not varied. Other
partial amplitudes having the spin-parity $J^\pi=1/2^-$, $1/2^+$ and $3/2^+$, as well as
a possible contribution in $J^\pi=3/2^-$ (in addition to $D_{13}(1520)$) were taken in
the form
\begin{equation}\label{tJP}
t_{J^P}=\left[t_B+t_R(W_{\gamma N})\right]G_{\Delta}F_{\Delta\to \pi N}\,.
\end{equation}
Here the first term in the brackets corresponds to the background, which is assumed to
depend smoothly on the energy, whereas the second term containing possible contributions
from baryon resonances in the $s$-channel, can undergo rapid variation with energy. The
factors $G_{\Delta}$ and $F_{\Delta\to \pi N}$ are the $\Delta$-isobar propagator and the
$\Delta\to\pi N$ decay vertex function. The details of parametrization of the terms $t_B$
and $t_R$ are described in Ref.\,\cite{KaFiPra}. The parameters were adjusted to the
angular distributions of the incident photons in the frame, in which the $z$-axis was
taken along the normal to the final state plane. The latter contains the momenta of all
three final particles in the center-of-mass system. The predictions of this fit for the
total cross section are presented in Fig.\,\ref{fig1} (the solid curve). In the same
figure we show the contributions coming from only the spin independent part $K$ and the
spin-flip amplitude $\vec{L}$.

The amplitude (\ref{tJP}) was fitted to the proton data. To calculate the photoproduction
on nuclei one needs information about the reaction on a neutron $\gamma n\to\pi^0\pi^0
n$. The contribution of $D_{13}(1520)$ to $\gamma n\to\pi^0\pi^0 n$ was determined by the PDG
values of the corresponding helicity $D_{13}\to n\gamma$ amplitudes $A_{1/2}$ and $A_{3/2}$ \cite{PDG}. To fix the
isotopic structure of other contributions we used the following simple prescription.
In the isotopic spin space of the nucleon the operator
$\hat{t}_{\gamma N}$ is given by
\begin{equation}\label{28}
\hat{t}_{\gamma N}=t_0+t_1\tau_3\,,
\end{equation}
where $\tau_3$ is the $z$-component of the nucleon isospin operator $\vec{\tau}$. The
coefficients $t_0$ and $t_1$ in Eq.\,(\ref{28}) determine transitions to the final
$\pi\pi N$ state when the initial nucleon absorbs a photon having isospin $I=0$ and
$I=1$, respectively. Therefore, the amplitudes of $\pi^0\pi^0$ photoproduction on the
proton and on the neutron read
\begin{equation}\label{32}
t_{\gamma p}=t_0+t_1\,, \quad t_{\gamma n}=t_0-t_1\,.
\end{equation}

A consequence of Eq.\,(\ref{32}) is that proton and neutron amplitudes are related to
each other as
\begin{equation}\label{36}
t_{\gamma n}=(2\alpha-1)\,t_{\gamma p}\,,
\end{equation}
where the parameter $\alpha$ is the ratio of the isoscalar part $t_0$ to the proton
amplitude
\begin{equation}\label{38}
\alpha=\frac{t_0}{t_{\gamma p}}\,.
\end{equation}
It determines the relative contribution of the isoscalar $(I=0)$ and the isovector
$(I=1)$ photons to the reaction. In the general case its value is energy dependent.
However for our qualitative estimations we took it as constant. To obtain $\alpha$ we
used experimental data for the total cross section of $\gamma n\to \pi^0\pi^0 n$,
presented in Ref.\,\cite{Ajaka2pi0n}. Since the cross section is proportional to the
square of the modulus of $t_{\gamma n}$, the value of $\alpha$ can not be unambiguously
determined. In particular, the data \cite{Ajaka2pi0n} for $\gamma n\to\pi^0\pi^0 n$ in Fig.\,\ref{fig1} may be described both with
$\alpha=0.08$ and $\alpha=0.75$. However, since the major part of the resonances in the energy
region considered are excited by the isovector photons and, therefore, the larger value
0.75 seems to be rather unlikely, we used
\begin{equation}\label{alpa}
\alpha=0.08\,.
\end{equation}

As is noted above, to calculate photoproduction of $\pi^0\pi^0$ on nuclei we used impulse
approximation. The kinematic in the $\gamma N$ subsystem was fixed using the prescription
of Ref.\,\cite{LaMar} in which the elementary amplitude $t_{\gamma N}$ is frozen at some
mean value of the initial nucleon momentum $\langle p_i\rangle$ determined in the overall center-of-mass
frame as
\begin{equation}\label{49}
\langle \vec{p}_i\rangle=-\frac{1}{A}\,\vec{k}-\frac{A-1}{2A}\,\vec{Q}\,,
\end{equation}
where $A$ is the nuclear mass number, $\vec{k}$ is the photon momentum, and $\vec{Q}$ is the three-momentum transferred to
the nucleus. The energy $E_i$ of the nucleon is fixed by the on-mass-shell condition
$E_i^2=M_N^2+\vec{p}^{\,2}_i$. The four-momentum $(E_f,\vec{p}_f)$ of the final nucleon
was then determined by the energy-momentum conservation at the single-nucleon vertex
\begin{eqnarray}
&&E_f=E_\gamma+E_i-\omega_1-\omega_2\,,\label{Ef}\\
&&\vec{p}_f=\vec{k}+\vec{p}_i-\vec{q}_1-\vec{q}_2\,,\label{pf}
\end{eqnarray}
where $(\omega_i,\vec{q}_i)$, $i=1,2$, is the four-momentum of the
$i$-th pion. The invariant $\gamma N$ energy $W_{\gamma N}$ in
(\ref{tJP}) was calculated as
\begin{equation}\label{WgN}
W_{\gamma N}=\sqrt{(E_\gamma+E_i)^2-(\vec{k}+\vec{p}_i\,)^2}\,.
\end{equation}
If the prescription (\ref{49}) together with the mass-shell condition for the initial
nucleon are used, the final nucleon turns out to be very close to its mass-shell. The
difference $\left|\,M_N-\sqrt{E_f^2-\vec{p}_f^{\ 2}}\,\right|$\,, where $E_f$ and
$\vec{p}_f$ are defined according to (\ref{Ef}) and (\ref{pf}), does not surpass 4 MeV in
the whole kinematic region considered. This is quite important in our case since our
amplitude (\ref{tJP}) is of pure phenomenological nature and is determined in a
relatively narrow energy region $W_{\gamma N}-(M_N+2M_\pi)\leq 325$\,MeV.
In Sect.\,\ref{Results} we also discuss sensitivity of
the results to the prescription for fixing the single nucleon kinematic.

\section{Results and discussion}\label{Results}

\begin{figure}
\begin{center}
\resizebox{0.8\textwidth}{!}{%
\includegraphics{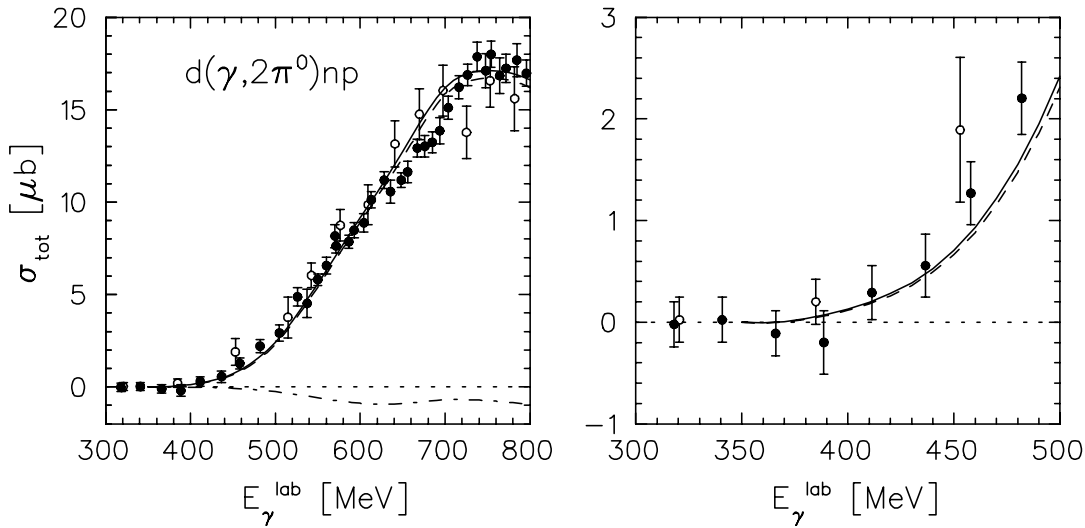}
} \caption{Total cross section of the incoherent reaction $\gamma d\to \pi^0\pi^0 np$.
The filled and open circles are taken from Ref.\,\cite{Kleber} and \cite{Krusche2pi0}
respectively. The dashed and the solid curves are obtained without and with inclusion of
the neutron-proton interaction in the final state. The dash-dotted curve represents the
value of the third term on the r.h.s.\ of Eq.\,(\ref{dnp}) (see discussion in the text). On the right
panel the low-energy region from threshold up to $E_\gamma=500$\, MeV is shown
separately.} \label{fig2}
\end{center}
\end{figure}

Using for $\alpha$ the value in (\ref{alpa}) we firstly calculated the cross section for
quasifree photoproduction of $\pi^0\pi^0$ on the deuteron. The results shown in
Fig.\,\ref{fig2} demonstrate rather good agreement with the data which is achieved within
only the impulse approximation, without much need for rescattering corrections. A direct
calculation of the nucleon-nucleon rescattering shows that this effect leads to a slight
increase of the total cross section in the whole energy region (compare dashed and solid
lines in Fig.\,\ref{fig2}). To include the neutron-proton interaction in the final state
we used the standard formalism described, for example, in Ref.\,\cite{FiAr2pi} and we
would like to refer the reader to this paper for more details.

One important point is in order. As is noted above, the incoherent cross section appears
to be rather insensitive to the model ingredients. For example, the dominant contribution
to incoherent photoproduction on a deuteron, provided by the impulse approximation, is
proportional to the incoherent sum $|t_{\gamma p}|^2+|t_{\gamma n}|^2$, folded with the
momentum distribution of the nucleons in the deuteron. As a consequence, if a realistic
nuclear wave function is used in conjunction with the elementary photoproduction operator
adjusted to the single nucleon data then already the impulse approximation provides a
rather good description of the unpolarized cross section on a nucleus. The mechanisms
involving more than one nucleon, like final state rescatterings, meson exchange currents,
nuclear isobar configurations {\it etc}, are important in rather narrow kinematical
region where the cross section is, as a rule, rather small. At the same time, in the
coherent channel the reaction amplitude is proportional to the absolute square of
$t_{\gamma p}+t_{\gamma n}$
\begin{equation}\label{42}
\sigma_{\gamma d}\sim |t_{\gamma p}+t_{\gamma n}|^2\,,
\end{equation}
so that not only the moduli but also the relative phase between $t_{\gamma p}$ and
$t_{\gamma n}$ play a role. The nature of this phase may be rather complicated. Even if
the reaction is dominated by the single resonance excitation, the nontrivial phase may
appear due to large background contributions in those channels to which the $\pi^0\pi^0$
channel is coupled through the unitary relations. This should primarily be the single
pion channel $\pi N$ which will generate a dressing of the bare vertex for resonance
photoexcitation $\gamma N\to R$, thus leading to an energy dependent electromagnetic
coupling $g_{\gamma N\to R}$ whose phase may be different for neutrons and protons. In
the case of $\eta$ photoproduction this question was addressed rather detailed in
Ref.\,\cite{Ritz}. In the present work we do not calculate these effects, since, as we
believe, their impact on our qualitative results is not crucial.

The amplitudes $t_{\gamma p}$ and $t_{\gamma n}$, obtained as described above, were then
used to calculate the cross sections of the coherent reactions
\begin{equation}\label{46}
\gamma+d\to\pi^0+\pi^0+d\,,
\end{equation}
\begin{equation}\label{48}
\gamma+^3\!\mbox{He}\to\pi^0+\pi^0+^3\!\mbox{He}\,.
\end{equation}

For the ground states of deuteron and $^3$He we adopted the phenomenological wave
functions from Refs.\,\cite{Mach} and \cite{Nisk}, respectively. These functions describe
the corresponding form factors in the whole momentum transfer region, as covered by the
energy region considered here. Since the deuteron has zero isospin, only the isoscalar
part $t_0$ in Eq.\,(\ref{28}) contributes to the reaction (\ref{46}). Using in
Eq.\,(\ref{15}) the deuteron quantum numbers and taking into account only the spherically
symmetric part in its wave function one obtains the approximate relation
\begin{equation}\label{52}
\sigma_{\gamma d}\sim
4\left[\langle|K_0|^2\rangle+\frac{2}{3}\langle|\vec{L}_0|^2\rangle\right]\,.
\end{equation}
Here the notation $\langle ... \rangle$ means averaging over the nuclear momentum
distribution, available for a given energy. The lower index ``0'' in $K_0$ and
$\vec{L}_0$ relates to the isoscalar part of these amplitudes in (\ref{24}).

\begin{figure}
\begin{center}
\resizebox{0.8\textwidth}{!}{%
\includegraphics{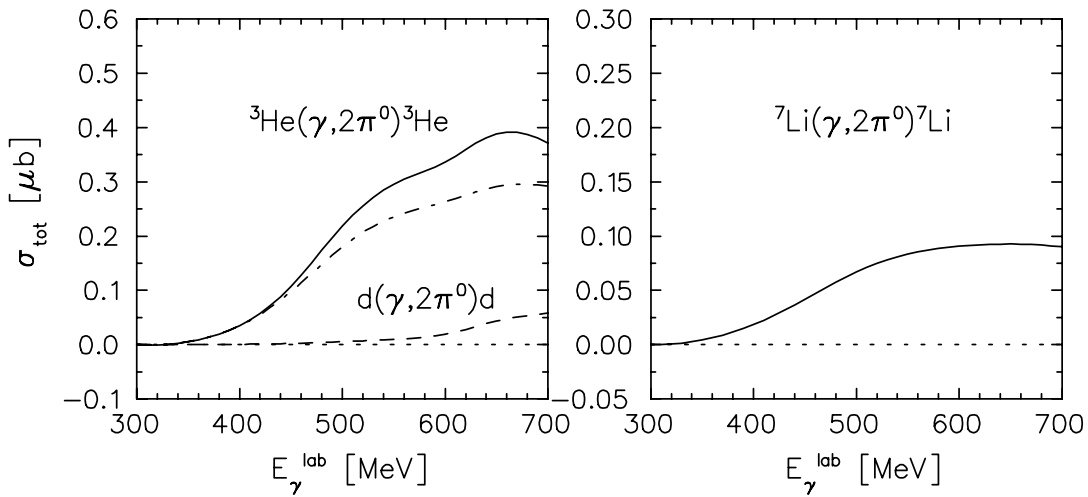}
} \caption{Left panel: total cross section for coherent $\pi^0\pi^0$ photoproduction on
the deuteron (dashed line) and $^3$He (solid line). The dash-dotted line is the cross
section on $^3$He, obtained with Blankenbecler-Sugar choice of the $\gamma N$ invariant
energy $W_{\gamma N}$ (see Eq.\,(\ref{BSW}) and the corresponding discussion in the
text). Right panel: total cross section for $\gamma\,^7$Li$\to\pi^0\pi^0\,^7$Li,
calculated with $^7$Li parameters from Ref.\,\protect\cite{Suesle}.} \label{fig3}
\end{center}
\end{figure}

According to the analysis outlined above the isoscalar part $t_0$ in Eq.\,(\ref{28})
amounts only 8$\%$ of the absolute value of the proton amplitude $t_{\gamma p}$.
Therefore, the cross section of the reaction (\ref{46}) is expected to be extremely
small. From the same considerations one can conclude that the cross section for $^3$He
should be at least one order of magnitude larger than that on the deuteron. Indeed, since
the isospin of $^3$He is 1/2, the corresponding cross section within the same
approximation as in Eq.\,(\ref{52}) reads
\begin{equation}\label{56}
\sigma_{\gamma ^3\!He}\sim
9\left[\langle|K_0+\frac13K_1|^2\rangle+\frac19\langle|L_0-L_1|^2\rangle\right]\,.
\end{equation}
Taking
\begin{equation}\label{58}
\left|\frac{K_0}{K_1}\right|\approx
\left|\frac{\vec{L}_0}{\vec{L}_1}\right|\approx8.7\cdot10^{-2}\,,
\end{equation}
according to (\ref{alpa}) and taking into account the approximate relation
\begin{equation}\label{60}
\big|K\big|\approx 1/\sqrt{2}\big|\vec{L}\big|\,,
\end{equation}
as follows from the calculations presented in Fig.\,\ref{fig1}, we obtain
\begin{equation}\label{62}
\frac{\sigma_{\gamma ^3\mbox{\scriptsize He}}}{\sigma_{\gamma d}}\approx 45\,.
\end{equation}
Our results for the cross sections (\ref{46}) and (\ref{48}) are plotted in
Fig.\,\ref{fig3}. As one can see, their ratio is in rough agreement with the estimation
(\ref{62}). At the energy $E_\gamma=450$\,MeV the value of $\sigma_{\gamma
^3\mbox{\scriptsize He}}/\sigma_{\gamma d}$ comprises about 46, at $E_\gamma=550$\,MeV it
reaches 31 and at $E_\gamma=700$\,MeV it decreases to 13. Thus, the cross section for
coherent $\pi^0\pi^0$ photoproduction should demonstrate a strong dependence on the
isospin of the target. This is primarily due to the smallness of the isoscalar part $t_0$
of the $\pi^0\pi^0$ photoproduction amplitude as is predicted by our calculation.

It is worth noting, that the smallness of the cross section for $\gamma d\to\pi^0\pi^0 d$
makes it possible to 'measure' the $\pi^0\pi^0$ production on a neutron as a difference
of the total inclusive cross sections on a deuteron and the free proton. Indeed, using
closure for the inclusive (coherent plus incoherent) cross section on a deuteron, one
obtains
\begin{equation}\label{dnp}
\sigma^{incl}_{\gamma d}\approx \langle\sigma_{\gamma p}\rangle+\langle\sigma_{\gamma
n}\rangle+2{\cal K}F_d(Q)Re \left\{\frac{1}{3}\langle\vec{L}^*_p\cdot \vec{L}_n\rangle
+\langle K^*_pK_n\rangle\right\}\,,
\end{equation}
where, as in the previous expressions, the notation $\langle ... \rangle$ means averaging
over the Fermi momentum distribution. ${\cal K}$ is a kinematical factor whose exact form
is insignificant for the present discussion. The last term, proportional to the deuteron
scalar matter form factor $F_d(Q)$, depending on the transferred three-momentum $Q$, is
generally rather small as may be seen from Fig.\,\ref{fig2}, where it is plotted by the
dash-dotted line.

It is also interesting to note one special property of photoproduction of two neutral
pions on $^3$He which is well seen in Eq.\,(\ref{56}). Namely, within the approximation,
in which only the spherically symmetric spatial part of the $^3$He nuclear wave function
is taken into account, the protons form a correlated pair with total spin $S=0$.
Therefore, the production of mesons with a nucleon spin-flip can proceed on the neutron
only. Otherwise we will have in the final state two protons with the same orbital and
spin momentum, contradicting the Pauli exclusion principle. This property is clearly
manifest in Eq.\,(\ref{56}), where the spin dependent part is proportional to
$|\vec{L}_0-\vec{L}_1|=|\vec{L}_{\gamma n}|$. This means, that one can use the ratio of
helium to proton cross sections as a measure of the nucleon spin-change probability in
$\pi^0\pi^0$ photoproduction.

Using the results presented in Fig.\,\ref{fig3} one can make a rough estimate of the
cross sections for $\pi^0\pi^0$ photoproduction on $^7$Li. Within a simple shell model
this nucleus is described as a system, containing a core, which is built up from four
nucleons with the total isospin $T=0$, and three valent nucleons. Because of smallness of
the isoscalar part $t_0$ of the amplitude $t_{\gamma N}$ the contribution of the core
should be small, and the major part of the cross section is expected to come from the
valent nucleons. In this respect the cross section for
$\gamma\,^7$Li$\to\pi^0\pi^0\,^7$Li may be estimated according to
\begin{equation}
\sigma_{\gamma^7\mbox{\scriptsize Li}}\approx R\ \sigma_{\gamma^3\mbox{\scriptsize He}}
\end{equation}
with
\begin{equation}
R=\frac{\int \Big[F_{^7\mbox{\scriptsize Li}}^{(1p)}(Q)\Big]^2d\Omega}{\int
\Big[F_{^3\mbox{\scriptsize He}}(Q)\Big]^2d\Omega}\,,
\end{equation}
where $Q$ is the momentum transfer, and $F_A(Q)$ are the corresponding nuclear matter
form factors. The notation $(1p)$ means that for $^7$Li only the contribution of the
$1p$-orbit is taken into account. The integration is performed over the phase space
available for each reaction. To estimate the ratio $R$ we used for a ground state $^7$Li
wave function a simple oscillator shell model, which gives for the form factors the well
known expressions:
\begin{equation}
F_{^3\mbox{\scriptsize He}}(Q)=\exp^{-(Qr_0)^2/4}\,,
\end{equation}
\begin{equation}
F_{^7\mbox{\scriptsize
Li}}^{(1p)}(Q)=\bigg(1-\frac{1}{6}(Qr_0)^2\bigg)\exp^{-(Qr_0)^2/4}\,.
\end{equation}
Taking $r_0=1.8$\,fm and $r_0=1.67$\,fm for $^3$He and $^7$Li, respectively, one obtains
for $R$ the value close to $0.41$ at $E_\gamma=400$\,MeV which decreases to $R=0.25$ at
$E_\gamma=750$\,MeV. In Fig.\,\ref{fig3} we present the cross section calculated with the
double-well $^7$Li wave function from Ref.\,\cite{Suesle}. The latter gives rather
satisfying description of the $^7$Li charge form factor in a wide range of momentum
transfer. As one can see, our results are in general agreement with the estimation $0.41
\leq R\leq 0.25$, obtained as discussed above.

Since the emphasis of our calculation for $\gamma\,^7$Li$\to\pi^0\pi^0\,^7$Li plotted in
Fig.\,\ref{fig3} was on demonstrating insignificance of the four-nucleon core, we do not
touch here upon some complications like, e.g., the role of the first excited $^7$Li level
$\frac12^-$ with $E^*=0.478$\,MeV. Since in the data, to which the $^7$Li wave function
in Ref.\,\cite{Suesle} was fitted, its contribution was subtracted, it also does not
enter our results. At the same time, due to obvious difficulty of separating this level
in $\pi^0\pi^0$ photoproduction, it should be taken into account if comparison of the
theoretical results with the data is intended.

Expression (\ref{28}) together with the smallness of the isoscalar part $t_0$ indicates
that the isotopic structure of $\pi^0\pi^0$ photoproduction is similar to that for $\eta$
photoproduction in the $S_{11}(1535)$ region. In the latter case, according to the
results of Refs.\,\cite{Krusche95,Sauermann,FiAr97,DoWe} the value of $\alpha$ is close
to 0.1. Due to the same reason the cross section for coherent $\pi^0\pi^0$
photoproduction on nuclei should have much in common with that for $\eta$
photoproduction. In particular, the process $A(\gamma,\eta)A$ is known to be suppressed
on the nuclei with zero isospin. For example, direct calculations within impulse
approximation as described in Refs.\,\cite{FiAr97} and \cite{FiArHe3}, give for the ratio
$\sigma_{\gamma\,^3\mbox{\scriptsize He}\to\eta\,^3\mbox{\scriptsize He}}/\sigma_{\gamma
d\to \eta d}$ in the region of $S_{11}(1535)$ the value about 20. The experimental
results for the $\gamma d\to\eta d$ and $\gamma\,^3$He$\to\eta\,^3$He cross sections
\cite{Krusche95,Pheron} indicate, that this ratio is likely to be essentially smaller,
about 4. This difference is primarily due to larger value of the deuteron cross section
in comparison to that predicted by the model of \cite{FiAr97} in which the possible
nontrivial phase between the neutron and the proton amplitudes is neglected (see
discussion after Eq.\,(\ref{42}) and in Ref.\,\cite{Ritz}).

Since our $\pi^0\pi^0$ amplitude has a relatively large resonance part $t_R$, as defined
in Eq.\,(\ref{tJP}), it should be sensitive to the variation of the $\gamma N$ invariant
energy $W_{\gamma N}$. As a consequence, the nuclear cross sections should strongly
depend on the prescription of choosing the $W_{\gamma N}$ value when the single-nucleon
operator is embedded into the nucleus. Quite apparently, this dependence brings
uncertainty into our calculations. To demonstrate the extent to which the results are
affected by this uncertainty we present in Fig.\,\ref{fig3} the cross section for
$\gamma\, ^3$He$\to\pi^0\pi^0\,^3$He which was calculated directly via integration over
the nucleon momentum $\vec{p}_i$ without resorting to the prescription (\ref{49}). The
value of $W_{\gamma N}$ was fixed using the so-called Blankenbecler-Sugar choice (see
also \cite{Breit} and \cite{FiArNPA}), corresponding to the assumption that all three
nucleons in the initial state of $^3$He share equally its energy, that is
\begin{equation}\label{BSW}
E_i^{lab}=\frac{1}{3}M_{^3\mbox{\scriptsize He}}\,,
\end{equation}
where $E_i^{lab}$ is the energy of the active nucleon in the $^3$He rest frame. The
corresponding result is shown on the left panel of Fig.\,\ref{fig3} by the dash-dotted
line. One readily notes that for coherent $\pi^0\pi^0$ photoproduction it is rather
important how the single nucleon kinematic is fixed. This is obviously a trivial
consequence of rather strong energy dependence of the elementary amplitude (\ref{tJP}).

Additional uncertainty comes from the assumption that $\alpha$ (\ref{38}) is energy
independent in a rather wide energy region. For the estimations presented in this paper
this point is perhaps not very crucial. However, better quantitative level of the
calculations requires more detailed knowledge of the isotopic structure of the elementary
amplitude. Obviously, for this task we need new data for $\pi^0\pi^0$ photoproduction on
a quasifree neutron with better than in Ref.\,\cite{Ajaka2pi0n} statistics, wherever
possible.

\section{Conclusion}
In this paper we present our predictions for coherent photoproduction of $\pi^0\pi^0$
pairs on the two lightest nuclei, $d$ and $^3$He. The calculations are based on the
available information about photoproduction on the nucleon, as well as on the physically
reasonable assumption, that in the second resonance region the $\pi^0\pi^0$ pairs are
photoproduced primarily by isovector photons. As our model shows, if these assumptions
are correct, the cross sections on protons and neutrons should be of an almost equal
size. In this case one should observe a relatively strong $\gamma\to \pi^0\pi^0$
transition on $^3$He and a rather weak transition on the deuteron. This situation is very
similar to that in $\eta$ photoproduction in the $S_{11}(1535)$ region, where the
isovector photons make the dominant contribution. One of the motivations for the present
paper was to encourage experimentalists to study the coherent $\pi^0\pi^0$
photoproduction on nuclei. The cross section of the process (\ref{48}) has appreciable
value, about 0.4 $\mu b$ and in principle can be measured with reasonable statistics.
Such measurements could be a good test of our knowledge about the spin-isospin structure
of the amplitude $\gamma N\to \pi^0\pi^0 N$.

\section*{Acknowledgment}
The authors acknowledge support from the Dynasty foundation, the TPU grant
LRU-FTI-123-2014 and the MSE program 'Nauka' (project 825)'.

\end{document}